\documentclass[10pt,journal]{IEEEtran}

\ifCLASSOPTIONcompsoc
\usepackage[nocompress]{cite}
\else
\usepackage{cite}
\fi
\ifCLASSINFOpdf
\else
\fi
\makeatletter
\def\ps@headings{%
	\def\@oddfoot{}%
	\def\@evenfoot{}}
\makeatother 

\usepackage[ruled,vlined,linesnumbered,ruled]{algorithm2e}
\usepackage{textcomp}
\usepackage{xcolor}
\usepackage{setspace}
\usepackage{siunitx}
\usepackage{makecell}
\usepackage{graphicx}
\usepackage{color}
\usepackage{ifpdf}
\usepackage{CJK}
\usepackage{amsfonts}
\usepackage{bbding}
\usepackage{booktabs}

\usepackage{cite}
\usepackage{multirow}
\usepackage{subfigure}
\usepackage{amssymb}
\usepackage{xspace}
\usepackage{epsf}
\usepackage{setspace}
\usepackage{tikz}
\usepackage{balance}
\usepackage{url,epsfig,array}
\usepackage{amsmath}
\usepackage{epstopdf}
\usepackage{cases}
\usepackage{bm}
\usepackage{amsthm}
\usepackage{enumitem}

\usepackage[flushleft]{threeparttable} 
\usepackage{etoolbox}      

\theoremstyle{definition}
\def\ie{\textit{i.e.}\xspace}

\def\etc{\textit{etc.}\xspace}

\def\eg{\textit{e.g.}\xspace}

\usepackage{supertabular}
\usepackage{pifont}

\usepackage{tikz}
\newcommand*{\circled}[1]{\lower.8ex\hbox{\tikz\draw (0pt, 0pt)%
    circle (.47em) node {\makebox[0.4em][c]{\small #1}};}}

\def\method{CoCoDC\xspace}

\begin{document}
\title{Cross-region Model Training with Communication-Computation Overlapping and Delay Compensation}
% \title{Training LLMs with Communication-Computation Overlapping and Delay Compensation}

\author{
    \IEEEauthorblockN{
    Ying Zhu,~Yang Xu,~Hongli Xu,~Yunming Liao,~Zhiwei Yao,~Liusheng Huang
    }\\
    \IEEEauthorblockA{
    School of Computer Science and Technology, University of Science and Technology of China\\
    Suzhou Institute for Advanced Research, University of Science and Technology of China\\
    \{isaaczhu, ymliao98, zhiweiyao\}@mail.ustc.edu.cn, \{xuyangcs, xuhongli, lshuang\}@ustc.edu.cn
    }
    % \thanks{* The Corresponding author.}
}

\IEEEtitleabstractindextext{%
\begin{abstract}
%The abstract goes here.
Training large language models (LLMs) requires massive computational resources, often necessitating the aggregation of geographically distributed datacenters (\ie, cross-region training). 
However, the high communication latency in wide-area networks severely degrades the efficiency of traditional distributed training. 
While methods like DiLoCo reduce communication frequency, they suffer from blocking synchronization.
Streaming DiLoCo alleviates this issue via communication-computation overlapping but introduces update staleness and model inconsistency due to delayed global updates and partial synchronization. 
These factors impair convergence, especially when aggressive overlap is needed to mask high latency.
We propose \method, a novel distributed training framework with communication-computation overlapping and delay compensation, to explicitly tackle such staleness and inconsistency. 
Within the \method framework, we specifically develop (1) a novel \textit{Delay Compensation} strategy based on Taylor expansion, and it approximates ideal model updates using local parameter and delayed global parameter changes, effectively mitigating the staleness; and (2) an \textit{Adaptive Transmission} strategy that dynamically schedules model fragment synchronization based on parameter changes and network capacity, optimizing bandwidth usage and accelerating convergence.
% Extensive experiments demonstrate that \method outperforms DiLoCo in terms of training efficiency and final model accuracy compared with Streaming DiLoCo, providing up to 21.0\% reduction in training steps required to reach similar perplexity.
Extensive experiments highlight the superior performance of \method over both DiLoCo and Streaming DiLoCo regarding final accuracy and training speed. Specifically, \method reduces the training steps needed to reach a comparable perplexity by up to 21.0\% compared to Streaming DiLoCo.
% , particularly under high network latency and aggressive overlap settings. 
Our work provides an effective solution for scalable and efficient cross-region LLM training.
\end{abstract}

\begin{IEEEkeywords}
%Computer Society, IEEEtran, journal, \LaTeX, paper, template.
% \emph{Decentralized Machine Learning, Edge Network, Device Heterogeneity, Peer Selection, Layer Selection}
\emph{Large Language Models, Cross-region Model Training, Communication-Computation Overlapping, Delay Compensation}
\end{IEEEkeywords}}

\maketitle

% is not selected <OR> if conference mode is selected - because compsoc
% conference papers position the abstract like regular (non-compsoc)
\IEEEdisplaynontitleabstractindextext

% \fi
\IEEEpeerreviewmaketitle

\ifCLASSOPTIONcompsoc
\IEEEraisesectionheading{\section{Introduction}\label{sec:introduction}}
\else
\section{Introduction}\label{sec:introduction}
\fi
The advent of large language models (LLMs) \cite{minaeelarge2025,artetxeEfficient2022,brownLanguage2020} has profoundly transformed the field of artificial intelligence (AI). 
These models, which often comprise billions or even trillions of parameters, have exhibited exceptional performance across a diverse range of natural language processing tasks.  
However, training such colossal models demands unprecedented computational resources, typically involving thousands of high-performance accelerators\cite{shoeybiMegatronLM2020,chowdheryPaLM2022} (\eg, GPUs or TPUs) operating continuously for weeks or even months. 
% Although scaling resources within a single datacenter  cluster has been the conventional approach, this strategy is increasingly constrained by challenges such as power delivery limitations, cooling infrastructure demands, and physical space restrictions. 
While scaling computational resources within single data center clusters remains the predominant strategy to train these LLMs, its feasibility is increasingly challenged by physical limitations, namely insufficient power delivery, cooling capacity constraints, and scarcity of physical space.

To leverage geographically distributed computational resources, training LLMs across multiple datacenters via wide-area networks (WANs) presents a compelling alternative\cite{Multidatacenter2024}. 
By pooling underutilized or idle resources from disparate locations, organizations can significantly reduce training durations, lower the substantial costs associated with dedicated clusters, and accelerate model development and iteration cycles. 
This distributed paradigm is especially pertinent in cross-region (or cross-datacenter) scenarios, such as federated learning (FL) \cite{mcmahancommunicationefficient2023,gorbettCrosssilo2024} across enterprise branches or collaborative research initiatives spanning multiple institutions, where data locality and privacy regulations may prevent centralized data aggregation.
However, a critical challenge in this distributed training framework is the high communication latency inherent in WANs\cite{liangCommunicationefficient2024}, which severely impedes traditional approaches like data parallelism (DP)\cite{shallueMeasuring2019}.
DP relies on synchronous stochastic gradient descent (SSGD) \cite{chenrevisiting2017}, requiring global gradient synchronization at each training step. This synchronization forces workers to remain idle during communication, leading to significant resource underutilization. 
The problem is further exacerbated in the context of LLMs, as their enormous parameter sizes substantially increase communication overhead, resulting in extended training delays.

To address the challenge of high communication overhead in cross-region environments, recent research has explored methods that reduce synchronization frequency.  
A notable approach is DiLoCo\cite{douillarddiloco2024,jaghouarOpenDiLoCo2024} (referred to as Distributed Low-Communication), where datacenters operate as workers, performing multiple local training steps (denoted by $H$) before synchronizing their local updates (\ie, pseudo-gradients) globally via an all-reduce operation. 
While DiLoCo effectively reduces synchronization frequency, it suffers from two primary limitations. 
Firstly, the global synchronization step forces all workers to halt local computation until communication completes.
This serialized execution of computation and communication inherently leads to severe resource underutilization, as computational resources remain idle during communication, while network bandwidth goes unused during computation, degrading overall wall-clock efficiency.
Secondly, despite the reduced frequency, each synchronization event in DiLoCo requires all workers to transmit their entire pseudo-gradients simultaneously.
This results in prohibitive peak bandwidth demands for environments constrained by low bandwidth or expensive inter-DC links.

To overcome these limitations of DiLoCo, Streaming DiLoCo \cite{douillardstreaming2025}, an advanced variant of DiLoCo, introduces optimizations: 
(a) Streaming Synchronization: The model parameters and their corresponding pseudo-gradients are partitioned along the depth dimension into smaller fragments.  
These fragments are then synchronized sequentially following a predefined schedule (\eg, evenly distributed across the $H$ local training), thereby smoothing peak bandwidth usage and avoiding the all-at-once transmission bottleneck.
(b) Communication-Computation Overlapping: 
Workers can concurrently perform local training on subsequent data batches while exchanging pseudo-gradient fragments from prior steps. 
This overlap, controlled by a parameter $\tau$ (called overlap depth), masks communication latency by enabling simultaneous model training and gradient transmission, thereby improving computational resource utilization and accelerating the overall training process.

However, these optimizations in Streaming DiLoCo introduce two new critical challenges: \textit{staleness} and \textit{inconsistency}\cite{narayananPipeDream2019,chenSAPipe2022}. 
Staleness arises from the temporal delay in gradient updates: when a worker receives and applies a global pseudo-gradient update for a specific model fragment, the pseudo-gradient is computed based on the global model state from $\tau$ steps earlier.
On the other hand, inconsistency stems from the partial nature of the updates inherent in the streaming paradigm. 
At any synchronization point, only the currently transmitted model fragment incorporates global information, while the remaining fragments retain their purely local state derived from the worker's independent computations. 
This creates an internal inconsistency across different parts of the model within a single worker. 
In general, both staleness and inconsistency represent deviations from the ideal synchronous update process. 
Staleness introduces errors due to outdated gradient information, while inconsistency creates misaligned model states within a single worker. 
Furthermore, when the overlap depth $\tau$ or the local training step $H$ increases, these deviations can accumulate, resulting in suboptimal parameter updates that hinder convergence speed and degrade final model performance.

To address the above issues in cross-region LLM training, we propose a novel distributed training framework with communication-computation overlapping and delay compensation, termed \method.
It introduces explicit strategies to mitigate the detrimental effects of staleness and inconsistency.
Specifically, we leverage Taylor expansion to approximate the ``ideal'' global pseudo-gradient information corresponding to the worker's current state, thereby compensating for the staleness introduced by the $\tau$ overlap steps.
Moreover, recognizing that model parameters contribute unevenly to convergence, we design an adaptive transmission algorithm that dynamically prioritizes fragments based on their importance or update magnitude.
This allows critical fragments to be transmitted multiple times within $H$ local training steps, optimizing bandwidth utilization and ensuring timely propagation of critical updates, rather than rigidly adhering to a fixed transmission schedule.
In summary, our main contributions are as follows: 
% 以下按照这三个方面进行组织：
% 1、针对DiLoCo里传算串行所导致的计算和通信资源利用不充分的问题，以及streaming DiLoCo里steleness和inconsistency所导致的xxx问题，我们提出了一个新的cross-region LLM training framework来通过delay compensation消除xxx的影响。
% 2、进一步对框架做个说明，并说明我们通过什么算法或策略来消除staleness和inconsistency的负面影响。
% 3、实验表明，我们相比于DiLoCo和Streaming DiLoCo的性能提升情况。

\begin{enumerate}

\item We identify critical limitations in existing cross-region methods: DiLoCo's serial execution leads to resource underutilization, while Streaming DiLoCo's overlap introduces staleness and inconsistency that can degrade convergence speed and final model performance. 
To address these, we propose \method, a novel distributed training framework specifically designed for efficient and robust cross-region LLM training.

\item \method employs a Taylor expansion-based delay compensation strategies to mitigate gradient staleness from communication overlap. 
Furthermore, it utilizes an adaptive block transmission strategy to optimize bandwidth and prioritize impactful parameter updates by transmitting them multiple times during the local training period, enhancing overall update quality and mitigating inconsistency effects.

\item Our extensive experiments validate that \method significantly reduces wall-clock training time compared to DiLoCo
% and DP 
by effectively overlapping communication and compensating for delay. 
Moreover, \method demonstrates superior convergence speed compared to Streaming DiLoCo, requiring up to 21.0\% fewer training steps to achieve comparable perplexity.
\end{enumerate}

% %\vspace{-1mm}
\section{Framework of \method}\label{sec:framework}
\subsection{DiLoCo and Streaming DiLoCo}

We consider the scenario of training an LLM collaboratively across $M$ geographically distributed datacenters under a decentralized, peer-to-peer (P2P) architecture. 
These datacenters, referred to as workers $W = \{w_1, w_2, ..., w_M\}$, are equipped with substantial computational resources, typically in the form of local accelerator clusters, and maintain distinct local data partitions $D_m$ ($m\in[1, M]$).
Notably, the data distributions across datacenters may be non-identically distributed (non-IID), reflecting realistic federated or collaborative learning settings.
Within each worker $w_m$, standard intra-node parallelism techniques, such as data parallelism (DP), tensor parallelism (TP), and pipeline parallelism (PP)\cite{narayananEfficient2021}, can be employed to optimize local training efficiency. 
However, these internal optimization mechanisms are orthogonal to the cross-region coordination protocol, which is the primary focus of this work.

The key challenge in the cross-region P2P training paradigm stems from the excessive communication overhead associated with synchronizing distributed model updates over WANs, which are typically characterized by high latency $L$ and limited bandwidth $B$. 
Traditional distributed training methods like standard DP require blocked gradient synchronization (\eg, ring all-reduce) after each local training step.
This leads to severe underutilization of computational resources and excessively prolonged training time. 

To mitigate this issue, DiLoCo reduces communication frequency by allowing each worker to perform $H$ local training steps before engaging in a global synchronization.
Rather than exchanging gradients, workers synchronize accumulated updates or pseudo-gradients, defined as the difference between their current local model $\theta^m_{t}$ and the globally agreed-upon model state from the previous synchronization round $\theta^g_{t-H}$. 
While this strategy effectively reduces the number of synchronization rounds, the global synchronization in DiLoCo remains a blocking operation, requiring all workers to wait for the completion of an all-reduce operation.
Moreover, even with infrequent synchronization, transmitting the entire model’s pseudo-gradients can still impose significant peak bandwidth demands.

To further enhance training efficiency, Streaming DiLoCo introduces two key optimizations: model partitioning and communication-computation overlapping. 
Specifically, the model parameters $\theta$ are partitioned along the depth dimension into $K$ disjoint fragments $\{\theta_1, ..., \theta_K\}$. 
Rather than synchronizing the entire model globally at intervals of $H$ steps, Streaming DiLoCo employs a fragment-wise synchronization strategy. This schedule distributes the synchronization of individual fragments across the $H$ local steps. 
The synchronization process proceeds as follows: 
At local step $t_p$, worker $w_m$ initiates the synchronization for fragment $p$ by computing the pseudo-gradient $\Delta \theta^m_p = \theta^m_{p, t_p} - \theta^g_{p, t_p-H}$, where $\theta^g_{p, t_p-H}$ represents the global state of fragment $p$ from its last synchronization $H$ steps earlier.
The worker then launches a non-blocking ring all-reduce operation for $\Delta \theta^m_p$, allowing local training to proceed concurrently.
While the communication happens in the background, the worker continues forward/backward passes on subsequent mini-batches. 
This overlap of computation and communication ensures that training progresses without waiting for synchronization to complete.

When communication for fragment $p$ completes after duration $\tau$, each worker has progressed to local step $t_l = t_p + \tau$. 
The overlap duration $\tau$, determined by the fragment size $S_p$ and network conditions, represents the effective communication delay experienced by workers for fragment synchronization. 
At step $t_l$, an outer optimizer (\eg, SGD with Nesterov momentum) uses $\Delta \theta^g_p$ to update the fragment's previous global state $\theta^g_{p, t_p-H}$ to compute the new target global state $\theta^{g}_{p, t_p}$:
\begin{align}
\label{eq:outer_grad_agg}
\Delta \theta^g_p &= \frac{1}{M} \sum_{m=1}^{M} (\theta^m_{p, t_p} - \theta^g_{p, t_p-H}) \\
\label{eq:outer_optim}
\theta^{g}_{p, t_p} &\leftarrow \text{OuterOptim}_p(\theta^g_{p, t_p-H}, \Delta \theta^g_p )
\end{align}
Streaming DiLoCo then blends the newly computed global state $\theta^{g}_{p, t_p}$ with the current local state $\theta^m_{p, t_l}$ of each worker $w_m$ using a mixing factor $\alpha$:
\begin{equation}
\label{eq:blending}
\theta^{m}_{p, t_l} \leftarrow (1-\alpha)\theta^{m}_{p, t_l} + \alpha \cdot \theta^{g}_{p, t_p}
\end{equation}
This blending operation incorporates global information into the local model while maintaining training progress.
However, as discussed in Section \ref{sec:introduction}, the communication-computation overlapping introduces staleness due to the $\tau$-step delay and inconsistency due to partial updates. 
The mixing factor $\alpha$ needs meticulous tuning to balance local and global updates.

\subsection{Our Method}

Our proposed \method directly addresses the above challenges through two key innovation strategies: \textit{delay compensation} and \textit{adaptive transmission}.
Instead of simply blending the stale global state, \method employs an explicit delay compensation strategy. 
% It leverages information available at the receiving step $t_l$, including the current local state $\theta^m_{p, t_l}$, the recorded local state $\theta^m_{p, t_p}$ from the time the synchronization started, and the received averaged pseudo-gradient $\Delta \theta^g_p$. 
The core challenge is that the information received at step $t_l$ reflects the global consensus relevant to step $t_p = t_l - \tau$. 
To apply a meaningful update at the current step $t_l$, we need to estimate how this global information should evolve over the delayed $\tau$ steps.
Mathematically, this involves approximating the ideal update at time $t_l$ based on information from $t_p$. 
Taylor expansion provides a natural framework for such extrapolation, approximating a function's future value based on its current value and derivatives \cite{liStragglers2021,bischofStructured1993}.
By estimating the ideal update at time $t_l$, we can effectively mitigates the negative impact of the staleness.
% Using a Taylor expansion-based approach}, %这里首先缺文献引用，另外这里引入的过于直白，为啥要用Taylor展开缺少必要的铺垫和说明。
% a \textit{corrected} update that approximates the ideal global update corresponding to the current time $t_l$ is computed and thus effectively mitigates the negative impact of the $\tau$-step staleness. 
Furthermore, to combat inconsistency and improve bandwidth utilization beyond the rigid scheduling of Streaming DiLoCo, \method incorporates an adaptive transmission strategy. 
This strategy estimates the maximum number of fragments that could be transmitted within the compute time of the $H$ local training steps.
% \rednote{during the computation time available within a synchronization cycle.} %这个可否简化一下描述？ 
It tracks the extent of parameter changes since the last synchronization. 
Instead of synchronizing each fragment exactly once per $H$ steps, \method prioritizes transmitting additional copies of more critical fragments within the available bandwidth and time window, aiming to propagate important updates faster and fill idle network capacity more effectively.

The core challenges in designing \method lie in formulating an effective and computationally feasible delay compensation strategy and developing a robust adaptive transmission strategy. 
We will elaborate on the strategy development in Section \ref{sec:strategy}.

\section{Optimization Strategies in \method}\label{sec:strategy}
In order to comprehensively address the challenges of resource underutilization, staleness, and inconsistency in cross-region distributed training environments, \method introduces two core optimization strategies :
1) communication and computation overlapping with delay compensation, and 
2) adaptive transmission to dynamically  schedule the synchronizations of pseudo-gradients fragments. %这里表述感觉有点问题 

\subsection{Delay Compensation}\label{subsec:delay_compensation}
Consider the training process for a specific worker $w_m$.
Upon completing the reception of the globally averaged pseudo-gradient $\Delta \theta^g_p$ for fragment $p$ at its local step $t_l$, this worker obtains information that reflects a consensus derived from states at an earlier step, specifically $t_p = t_l - \tau$. 
Directly applying this potentially stale information, for instance, through a simple blending mechanism as described in Eq. \eqref{eq:blending}, can lead to inaccurate model updates. 
This issue becomes particularly pronounced as the scale of the LLM increases and/or the overlap depth $\tau$ gets large.

Consequently, the objective of the delay compensation strategy within \method is to refine the update applied at step $t_l$.
This refinement involves estimating how the ideal global update would likely have evolved during the $\tau$ steps of overlapping.
The goal is to compute a corrected target state $\theta^m_{p,t_l}$, which more accurately approximates the desired global consensus state pertinent to the current time step $t_l$ rather than the stale time step $t_p$.
Drawing inspiration from prior research in asynchronous SGD methodologies \cite{zhengAsynchronous2020, rigazziDCS3GD2019, wangAsynchronous2023}, we employ Taylor expansion to predict the gradient at the current time step. 
However, it is crucial to note that these preceding methods primarily address training scenarios involving relatively smaller models within a single DC,  typically transmitting instantaneous gradients after each local step (\ie, $H=1$) where $\tau$ is comparatively small.
In contrast, \method involves the transmission of pseudo-gradients computed by accumulating model updates over a potentially large number of local steps (\ie, $H \gg 1$) and incorporates compensation when $\tau > 1$.
Therefore, the aforementioned prior methods can be viewed as representing simpler, specialized cases within the more general framework of \method.

Let $\theta^g_{p,t_p}$ denote the reference global state for fragment $p$ computed at step $t_l$ after applying the outer optimization to the state $\theta^g_{p, t_p-H}$ using the received $\Delta \theta^g_p$ as in Eq. \eqref{eq:outer_optim}. 
Thus, it represents the global consensus state at step $t_p$. 
Consider the ideal but unknown evolution of the global parameters for fragment $p$. 
A first-order Taylor expansion of the parameters around time $t_p$ could be written conceptually as:
$$\theta^g_{p, t_l} \approx \theta^{g}_{p, t_p} + \nabla_{\theta} \theta^g_{p,t_p} \cdot \tau$$
where $\nabla^m_{\theta} \theta^g_{p,t_p}$ represents the true instantaneous rate of change of the ideal global parameters at time $t_p$.
% \rednote{Since we cannot directly compute this, we leverage locally available information.
% The worker $w_m$ knows its own parameters at the start of the overlap, $\theta^m_{p, t_p}$, and at the end, $\theta^m_{p, t_l}$. }
Although the worker will obtain $\theta^{g}_{p, t_p}$ after synchronization, it is inefficient to compute $\nabla_{\theta} \theta^g_{p,t_p}$ using the same data batches again.
Thus we compute the average change rate of the worker's local parameters during the $\tau$ overlap steps as an estimate of $\theta^{g}_{p, t_p}$:
% As an estimate for the update \rednote{velocity}, we compute the average velocity of the worker's local parameters during the \rednote{overlap period}:
\begin{equation}
\label{eq:local_velocity_formal}
g^m_{p,t_p} = \frac{\theta^m_{p, t_p} - \theta^m_{p, t_l}}{\tau}
\end{equation}
% This local velocity $g^m_{p,t_p}$ serves as a first approximation of the desired update direction. 
However, $\theta^{g}_{p, t_p}$ is biased by the worker's local data, and the starting point $\theta^m_{p, t_p}$ has diverged from the global state.
We apply Taylor expansion's principle to the change rate around the global parameter $\theta^g_{p,t_p}$:
\begin{align}
g^{ideal,m}_{p,t_p} &\approx g^m_{p,t_p} + \nabla g^m_{p,t_p}\cdot\frac{(\theta^g_{p, t_p} - \theta^m_{p, t_p})}{H} \\
&= g^m_{p,t_p} + H^m_{p,t_p}\cdot \frac{\Delta \theta^m_{p,t_p}}{H} 
\end{align}
where $H^m_{p,t_p}$ is the Hessian of $g^m_{p,t_p}$ at $\theta^m_{p,t_p}$.
Note that the model difference $\Delta \theta^m_{p,t_p}$ is divided by $H$ to account for the fact that it is accumulated over the previous $H$ steps and we need to align it with the scale of other terms.

Since directly computing the Hessian is intractable for LLMs since it takes $O(n^2)$ storage for $O(n)$ parameters (\eg, a trillion-parameter model would require exabytes for the full Hessian),
we approximate the action of the Hessian using the outer product of the local parameters' change rate.
Specifically, the term $H^m_{p,\cdot}$ is approximated by $\lambda \cdot g^m_{p,\cdot} \odot g^m_{p,\cdot}$, where $\lambda$ is a tunable hyperparameter scaling the compensation strength and $\odot$ denotes the outer product operation.
This is justified by the connection between the gradient outer product and the Fisher Information Matrix\cite{friedman2001elements,amariNatural1998}.
Substituting the approximated Hessian into the conceptual change rate correction:
\begin{equation}
\label{eq:corrected_velocity_formal}
g^{corr,m}_{p,t_p} = g^m_{p,t_p} + \lambda\cdot g^m_{p,t_p} \odot g^m_{p,t_p} \odot \frac{\Delta \theta^m_{p,t_p}}{H}    
\end{equation}
This $g^{corr,m}_{p,t_p}$ represents the estimate of the appropriate average change rate for fragment $p$ during the interval $[t_p, t_l]$, adjusted for the observed divergence between the local model and the global updates.

Finally, we compute the corrected target state $\theta^{m}_{p,t_l}$ by applying this corrected change rate over the $\tau$ steps, starting from the updated global state $\theta^{g}_{p,t_p}$:
\begin{equation}
\label{eq:corrected_target_formal}
\theta^{m}_{p,t_l} \leftarrow \theta^{g}_{p,t_p} + g^{corr,m}_{p,t_p} \times \tau
\end{equation}
After the above compensation steps, the worker $w_m$ can continue training on the updated $\theta^{m}_{p,t_l}$.
The whole compensation algorithm is summarized in Algorithm \ref{alg:delay_compensation_formal}.

\begin{algorithm}[t]
\caption{Delay Compensation for worker $w_m$ in \method}
\label{alg:delay_compensation_formal}
\KwIn{Local parameters $\theta^m_{p, t_l}$, $\theta^m_{p, t_p}$; global state $\theta^{g}_{p, t_p}$; overlap depth $\tau$; local computation period length $H$; compensation strength $\lambda$}
\KwOut{Updated local parameters $\theta^m_{p, t_l}$}
Calculate $g^m_{p,t_p}$ during $\tau$ using Eq. \eqref{eq:local_velocity_formal}\;
Estimate correct $g^{corr,m}_{p,t_p}$ using Eq. \eqref{eq:corrected_velocity_formal}\;
Obtain $\theta^m_{p, t_l}$ by applying $g^{corr,m}_{p,t_p}$ to global state $\theta^m_{p, t_p}$ using Eq. \eqref{eq:corrected_target_formal}\;
\Return{$\theta^m_{p, t_l}$}
\end{algorithm}

% % 3.2 Adaptive Transmission Strategy
\subsection{Adaptive Transmission}\label{subsec:adaptive_transmission}

The fixed, round-robin fragment synchronization schedule employed in Streaming DiLoCo, where each fragment $p$ is synchronized exactly once every $H$ steps, exhibits some limitations in dynamic cross-region environments. 
Firstly, when the actual fragment synchronization time $T_s$ (primarily dominated by the latency induced by all-reduce operation over WANs) 
is considerably shorter than the allocated time per fragment based on local computation speed ($T_c \times H/K$), considerable network bandwidth remains unutilized during training.
Secondly, the rigid synchronization schedule negelects the learning dynamics during training, as some fragments might undergo more substantial updates and would benefit from more frequent synchronization to propagate critical updates faster, while relatively stable fragments could tolerate reduced synchronization frequency. 

Our adaptive transmission strategy aims to address these shortcomings by transmitting rapidly changing fragments more frequently while still preventing any fragment from becoming excessively stale, which contributes to more effective utilization of the available bandwidth. 
This strategy operates continuously, making decisions whenever a worker is ready to initiate a new fragment synchronization.

The implementation begins with estimating the average time per local computation step $T_c$ and the average single-fragment synchronization time $T_s$. 
These values are obtained either through initial benchmarking or via online moving averages during training.  
To ensure robustness against network variability and prevent detrimental network congestion, we introduce a network utilization factor $\gamma \in (0, 1]$ to modulate scheduling aggressiveness. 
% This factor moderates the aggressiveness of the adaptive scheduling. 
The target number of synchronizations ($N$) achievable within $H$ local training steps is estimated as:
\begin{equation}
\label{eq:target_syncs}
N = \max\left(K, \lfloor \gamma \times \frac{H \times T_c}{T_s} \rfloor\right)
\end{equation}
This formulation guarantees that at least all $K$ fragments are synchronized once per $H$ steps, maintaining the baseline behavior as a minimum while permitting additional synchronizations up to available capacity. 
Consequently, the average synchronization interval $h$ between initiating consecutive transmissions becomes:
\begin{equation}
\label{eq:effective_interval}
h = \lfloor H / N \rfloor
\end{equation}

To guide the adaptive fragment selection process, we introduce a quantitative metric that captures the relative importance of each fragment's updates.
This metric leverages the globally averaged pseudo-gradients $\Delta \theta^g_p$ received by each worker $w_m$ upon completing fragment synchronization at global step $t_l$.
% Since each worker $w_m$ receives the globally averaged pseudo-gradient $\Delta \theta^g_p$ for fragment $p$ upon completion of its synchronization at global step $t_l$, we can leverage this information. 
Let $t_{p,b}$ denote the global step at which the previous synchronization for fragment $p$ was completed. 
Then, the interval duration is $I_p = t_p - t_{p,b}$, and we define a metric $R_{p, t_p}$ representing the average change rate for parameters in fragment $p$:
\begin{equation}
\label{eq:impact_metric}
R_{p, t_p} = \|\Delta \theta^g_p\|_2 / I_p
\end{equation}
A higher $R_p$ suggests the fragment is undergoing significant adjustments based on the collective progress. 
This metric is updated for fragment $p$ whenever its synchronization completes at step $t_l$. 
For fragments that have not yet completed a synchronization, $R_p$ is initialized to infinity to ensure initial transmission priority.

The core decision logic runs whenever a worker determines it is ready to initiate the next fragment synchronization. 
Let the current global step be $t_{current}$.
% \rednote{We first identify any fragment $p$ whose time since its last synchronization $t_{current} - t_{p,b}$ is approaching or exceeding the maximum allowed interval $H$.}
We first identify if any fragment $p$ has not been synchronized for at least $H$ training steps, \ie, $I_p \geq H$.
If such a fragment exists, it will be selected as $p^*$; else the fragment $p^*$ that maximizes the impact metric $R_p$ among all fragments $p \in \{1, ..., K\}$ will be selected.
\begin{equation}
\label{eq:fragment_selection}
p^* = argmax_{p \in \{1, ..., K\}}R_p
\end{equation}

This selection process is deterministic since all workers have access to the same history of globally averaged pseudo-gradients $\Delta \theta^g_p$. 
This eliminates the need for explicit coordination messages for the selection itself. 
The selected fragment $p^*$ is then prepared for synchronization, calculating its pseudo-gradient $\Delta \theta^m_{p^*}$ for the upcoming all-reduce operation, \etc
The complete adaptive selection procedure is formally presented in Algorithm \ref{alg:adaptive_selection}.

\begin{algorithm}[t]
% \caption{\rednote{Adaptive Transmission Strategy - Fragment Selection}} %是否有更简洁的标题
\caption{Fragment Selection in \method}
\label{alg:adaptive_selection}
\KwIn{$t_{g}$, $H$, $K$, $R_p$ for each $p$}
\KwOut{The next fragment to synchronize, $p^*$}
% \tcp{Check for excessive staleness}
\tcp{Check for fragments that aren't synchronized for a long period}
\For{each fragment $p \in \{1, ..., K\}$}{
    \If{$(t_{g} - t_{p,b}) \ge H$}{
        $p^* \leftarrow p$\; 
        \Return{$p^*$}
    }
}
\tcp{Select fragment with the largest $R$}
$p^* \leftarrow \arg\max_{p \in \{1, ..., K\}} R_p$\;
\Return{$p^*$}
\end{algorithm}

% This adaptive strategy offers several advantages.
% By allowing more than $K$ synchronizations within an $H$-step cycle when capacity permits ($N \geq K$), it can fill idle network time and improve network utilization.
% The utilization factor $\gamma$ provides a buffer against network fluctuations or inaccurate time estimations.
% It also focuses communication bandwidth on fragments demonstrating larger global impact, potentially accelerating convergence for critical model parts.
% The mandatory synchronization ensures no fragment lags behind by more than $H$ steps.
% All workers use the same historical global information and the same deterministic rules to select the next fragment, eliminating the need for complex voting or extra communication for decision-making.

\section{Experiments and Evaluation}\label{sec:evaluation}
% In this section, we compare \method against relevant baselines to validate its effectiveness in mitigating stalenss and inconsistency introduced in cross-region LLM training. 

\begin{figure}[t]
    \centering
    \includegraphics[width=0.85\linewidth]{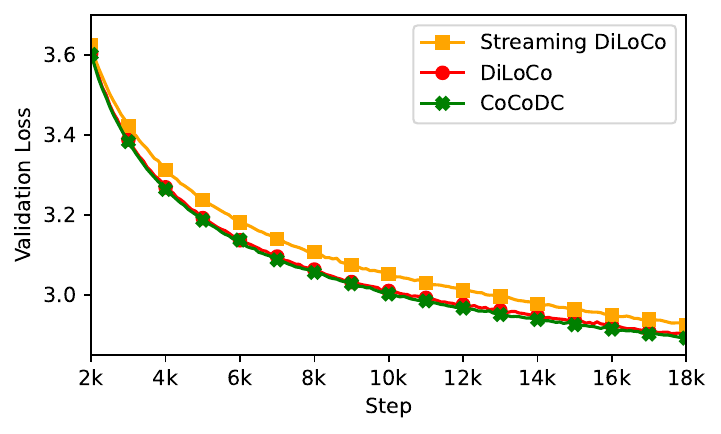}
    \caption{Validation loss vs. training steps of different methods}
    \label{fig:val_loss_comp}
\end{figure}

\begin{figure}[t]
    \centering
    \includegraphics[width=0.85\linewidth]{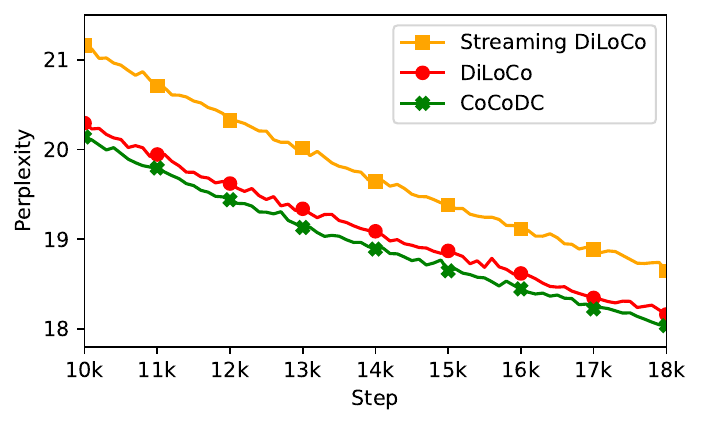}
    \caption{Validation perplexity vs. training steps of different methods}
    \label{fig:val_ppl_comp}
\end{figure}

% \begin{table}[ht] % 'ht' suggests placement here or top
%     \centering % Center the table horizontally
%     \caption{Comparison of Final Validation Metrics and Convergence Speed for Different Methods}
%     \label{tab:comparison_results} % Label for referencing
%     \begin{tabular}{lccc}
%         \toprule % Top rule from booktabs
%         \textbf{Method} & \textbf{Final Val Loss} & \textbf{Final Val Perplexity} & \textbf{Steps (to PPL $\le$ 20.0)} \\
%         \midrule % Middle rule from booktabs
%         \method{}           & 2.8924 & 18.0357 & 10,292 \\ % Use your final method name
%         Streaming DiLoCo    & 2.9256 & 18.6457 & 13,021 \\
%         DiLoCo    & 2.8993 & 18.1608 & 10,821 \\
%         \bottomrule % Bottom rule from booktabs
%     \end{tabular}
% \end{table}

% \begin{table}[ht]
%     \centering
%     \caption{Comparison of Final Validation Metrics and Convergence Speed for Different Methods}
%     \label{tab:comparison_results_abbr}
%     \begin{tabular}{S[table-format=1.0] S[table-format=1.2] S[table-format=2.4] S[table-format=5.0]} 
%         \toprule
%         \textbf{Method} & {\textbf{Validation}} & {\textbf{Validation}} & {\textbf{Steps}} \\ % Abbreviated headers
%                         &     \textbf{Loss}         & \textbf{Perplexity}                   & {\textbf{(PPL $\le$ 20)}} \\ % Second line for Steps condition
%         \midrule
%         CoCoDC           & 2.8924 & 18.0357 & 10292 \\
%         Streaming DiLoCo & 2.9256 & 18.6457 & 13021 \\
%         DiLoCo           & 2.8993 & 18.1608 & 10821 \\
%         \bottomrule
%     \end{tabular}
% \end{table}
\begin{table}[ht]
    \centering
    \sisetup{detect-weight=true, detect-family=true} 
    \robustify\bfseries 
    \begin{threeparttable}
        \caption{Comparison of Final Validation Metrics and Convergence Speed for Different Methods}
        \label{tab:comparison_results_revised}
        % \vspace{2mm} 
        \begin{tabular}{c % Method 列：改为 c 实现居中对齐
                        S[table-format=1.4] % Val Loss 列
                        S[table-format=2.4] % Val PPL 列
                        S[table-format=5.0]} % Steps 列
            \toprule
            % 第一行表头：使用 \multirow 使前三列跨两行并垂直居中
            % Steps 表头只在第一行
            \multirow{2}{*}{\textbf{Method}} & 
            {\multirow{2}{*}{\textbf{Loss}}} & 
            {\multirow{2}{*}{\textbf{PPL}}} & 
            {\textbf{Steps}} \\ 
            % 第二行表头：前三列留空，因为它们被 \multirow 占据
            % 只有 Steps 列的第二部分内容
             & & & {\textbf{(PPL $\le$ 20)}} \\ 
            \midrule
            Streaming DiLoCo & 2.9256 & 18.6457 & 13021 \\
            DiLoCo           & 2.8993 & 18.1608 & 10821 \\
            CoCoDC           & \textbf{2.8924} & \textbf{18.0357} & \textbf{10292} \\ 
            \bottomrule
        \end{tabular}
        \begin{tablenotes}[para,flushleft] % para 选项使注释呈段落状，flushleft 左对齐
            \small 
            \item[] \textbf{PPL}: Perplexity. \textbf{Steps} column shows the number of steps required to reach a Perplexity of 20.0.
        \end{tablenotes}
    \end{threeparttable}
\end{table}
% \begin{table}[ht]
%     \centering
%     \caption{Comparison of Final Validation Metrics and Convergence Speed for Different Methods}
%     \label{tab:comparison_results_revised}
%     \begin{tabular}{l % Method name (left aligned text)
%                     S[table-format=1.4] % Val Loss (1 digit before, 4 after decimal)
%                     S[table-format=2.4] % Val PPL (2 digits before, 4 after decimal)
%                     S[table-format=5.0]} % Steps (5 digits before, 0 after decimal)
%         \toprule
%         \textbf{Method} & {\textbf{Val Loss}} & {\textbf{Val PPL}} & {\textbf{Steps}} \\
%                         &                     &                    & {\textbf{(PPL $\le$ 20)}} \\ 
%         \midrule
%         CoCoDC           & 2.8924 & 18.0357 & 10292 \\ 
%         Streaming DiLoCo & 2.9256 & 18.6457 & 13021 \\
%         DiLoCo           & 2.8993 & 18.1608 & 10821 \\
%         \bottomrule
%     \end{tabular}
%     \textbf{PPL}: Perplexity. \textbf{Steps} column shows the number of steps required to reach a Perplexity of 20.0.
% \end{table}

\subsection{Experimental Settings}
% \textbf{Environment:} 
We simulate a cross-region distributed training environment consisting of $M=4$ workers, representing geographically distinct datacenters. 
The experiments are conducted on a single server equipped with 4 NVIDIA A100 GPUs (80GB memory), where each GPU emulates one worker node. 
Inter-worker communication is modeled using a ring all-reduce topology implemented through PyTorch's distributed training framework \cite{liPyTorch2020}.
% For the results presented, the computational throughput of all workers is assumed to be homogeneous.
We assume homogeneous computational capacities across all workers to isolate the effects of our proposed optimizations.

\textbf{Dataset and model:} We employ a decoder-only and LLaMA-style transformer\cite{touvronLLaMA2023,jaghouarOpenDiLoCo2024}, configured with 12 decoder layers and approximately 150 million parameters.  
The model is trained on the English portion of the C4 dataset (a colossal, cleaned version of Common Crawl's web crawl corpus) \cite{raffelExploring2023} for a language modeling task.

\textbf{Training Hyperparameters:}
All models are trained for a total of 18,000 local training steps.
For local training optimizer, we use AdamW with a learning rate of $4 \times 10^{-4}$, and a weight decay of 0.1.
A linear warmup phase of 1000 steps is followed by a cosine learning rate decay schedule. 
We utilize Automatic Mixed Precision (AMP) with bf16 format\cite{kalamkarStudy2019} for computational efficiency, and use high precision for matrix multiplications for accuracy.
We employ a global effective batch size of 256 sequences per optimizer step with a sequence length of 1024.

\textbf{Baselines and Metrics:}
We compare \method against DiLoCo and Streaming DiLoCo. 
For all methods, the number of local computation steps $H$ is set to 100.
To simulate network constraints, the overlap depth $\tau$ is set to 5 for Streaming DiLoCo and our \method.
The fragmentation method are the same in \method and Streaming DiLoCo, with approximately 3 layers per shard using a strided pattern as in Streaming DiLoCo, thus formulating a total of $4$ shards.
For \method, we use a compensation factor $\lambda$ of 0.5 and network utilization factor $\gamma=0.4$, resulting in 8 synchronizations per $H$ steps.
Performance is evaluated periodically during training by calculating the loss and perplexity on a held-out validation split of the C4 dataset.

\subsection{Results and Analysis}

Figs \ref{fig:val_loss_comp} and \ref{fig:val_ppl_comp} present the comparative performance of \method against baseline methods.
The results demonstrate that \method consistently maintains lower validation loss and perplexity than both Streaming DiLoCo and DiLoCo throughout the training process.
Specifically, as indicated in Table \ref{tab:comparison_results_revised}, \method achieves a perplexity of 20.0 while requiring 21.0\% fewer training steps (10292) than Streaming DiLoCo (13021) and 4.9\% fewer steps than DiLoCo (10821) to reach comparable performance levels.
\method also achieves the lowest perplexity on the validation set among all the methods.
While DiLoCo eventually attains similar convergence metrics to \method, this comes at the cost of significantly longer training time for synchronizing pseudo-gradients of the full model.
In contrast, \method delivers comparable (and often superior) convergence characteristics while demonstrating substantially improved wall-clock time efficiency.
While the current experiments may not show substantial improvements in final metrics over baseline methods, the advantages of delay compensation and adaptive transmission are expected to become significantly more pronounced under more aggressive, real-world cross-region conditions characterized by much higher network latencies.
Such conditions would necessitate larger $H$ and $\tau$ to hide synchronization latency and amortize synchronization overhead, consequently magnifying the staleness and inconsistency effect that \method is designed to counteract. 
We plan to conduct more comprehensive evaluations, including larger model architectures, more diverse datasets, and real-world deployment scenarios with large $H$ and $\tau$, and present relevant experimental results in future work.
% Experiments under these aggressive settings will further validate the robustness and generalizability of our method.

The experimental results reveal several key insights. 
Firstly, the communication-computation overlapping in Streaming DiLoCo introduces measurable performance degradation due to update staleness. 
Secondly, the compensation strategy in \method effectively mitigates this negative effect, yielding both faster convergence and better final model performance. 
The superior performance of \method compared to Streaming DiLoCo with delay compensation under identical overlap conditions confirms the importance of explicit staleness handling. 
Furthermore, the ability of \method to outperform even synchronous approaches like DiLoCo suggests that carefully managed frequent synchronization can enhance model convergence beyond what traditional synchronous methods achieve.
% While the current experiments demonstrate the effectiveness of our approach under some controlled conditions ($H=100,\tau=5$), we plan to conduct more comprehensive evaluations in future work. 
% This will include larger-scale experiments with larger model architectures, more diverse datasets, and real-world deployment scenarios to further validate the generalizability of our method.

\section{Conclusions}\label{sec:conclusion}
% In this paper, we have proposed a heterogeneity-aware FML framework, termed \method, and developed an algorithm to dynamically and adaptively adjust the inner-update frequency and meta-update frequency for heterogeneous clients to accelerate model training in edge networks.
% We have conducted extensive experiments on a hardware prototype involving 50 NVIDIA Jetson devices.
% The experimental results have demonstrated that \method can not only achieve a speedup of model training by up to 4.02$\times$ but also improve model accuracy by an average of 11.5\% on evaluation clients compared to the baselines.

% ---
% In this work, we proposed \method, a novel distributed framework for efficient cross-region LLM training. 
% \method utilizes a delay compensation technique derived from Taylor approximation and an adaptive transmission strategy to dynamically manage fragment synchronization. 
% This approach harnesses the benefits of communication-computation overlapping while counteracting the effects of staleness. 
% Our experimental resutls demonstrate that \method outperforms both Streaming DiLoCo and DiLoCo, achieving lower validation loss and perplexity per step. 
% \method also reduced the steps needed to reach a target perplexity by up to 21.0\% compared to Streaming DiLoCo, underscoring the effectiveness of our optimization strategies.

In this work, we present \method, a novel distributed framework for efficient cross-region LLM training. 
By integrating Taylor-based delay compensation and adaptive fragment synchronization, \method effectively leverages communication-computation overlapping while mitigating staleness. 
Experiments show that \method outperforms both Streaming DiLoCo and DiLoCo, achieving lower validation loss and perplexity per step and reducing the steps to target perplexity by up to 21.0\%. 
These results validate the efficacy of our optimization strategies.

\balance
\bibliographystyle{IEEEtran}
\bibliography{contents/refs}

\begin{thebibliography}{10}
\providecommand{\url}[1]{#1}
\csname url@samestyle\endcsname
\providecommand{\newblock}{\relax}
\providecommand{\bibinfo}[2]{#2}
\providecommand{\BIBentrySTDinterwordspacing}{\spaceskip=0pt\relax}
\providecommand{\BIBentryALTinterwordstretchfactor}{4}
\providecommand{\BIBentryALTinterwordspacing}{\spaceskip=\fontdimen2\font plus
\BIBentryALTinterwordstretchfactor\fontdimen3\font minus \fontdimen4\font\relax}
\providecommand{\BIBforeignlanguage}[2]{{%
\expandafter\ifx\csname l@#1\endcsname\relax
\typeout{** WARNING: IEEEtran.bst: No hyphenation pattern has been}%
\typeout{** loaded for the language `#1'. Using the pattern for}%
\typeout{** the default language instead.}%
\else
\language=\csname l@#1\endcsname
\fi
#2}}
\providecommand{\BIBdecl}{\relax}
\BIBdecl

\bibitem{minaeelarge2025}
\BIBentryALTinterwordspacing
S.~Minaee, T.~Mikolov, N.~Nikzad, M.~Chenaghlu, R.~Socher, X.~Amatriain, and J.~Gao. \BIBforeignlanguage{en}{{Large Language Models: A Survey}}. [Online]. Available: \url{http://arxiv.org/abs/2402.06196}
\BIBentrySTDinterwordspacing

\bibitem{artetxeEfficient2022}
M.~Artetxe, S.~Bhosale, N.~Goyal, T.~Mihaylov, M.~Ott, S.~Shleifer, X.~V. Lin, J.~Du, S.~Iyer, R.~Pasunuru, G.~Anantharaman, X.~Li, S.~Chen, H.~Akin, M.~Baines, L.~Martin, X.~Zhou, P.~S. Koura, B.~O'Horo, J.~Wang, L.~Zettlemoyer, M.~Diab, Z.~Kozareva, and V.~Stoyanov, ``\BIBforeignlanguage{en}{{Efficient Large Scale Language Modeling with Mixtures of Experts}},'' Oct. 2022.

\bibitem{brownLanguage2020}
T.~B. Brown, B.~Mann, N.~Ryder, M.~Subbiah, J.~Kaplan, P.~Dhariwal, A.~Neelakantan, P.~Shyam, G.~Sastry, A.~Askell, S.~Agarwal, A.~{Herbert-Voss}, G.~Krueger, T.~Henighan, R.~Child, A.~Ramesh, D.~M. Ziegler, J.~Wu, C.~Winter, C.~Hesse, M.~Chen, E.~Sigler, M.~Litwin, S.~Gray, B.~Chess, J.~Clark, C.~Berner, S.~McCandlish, A.~Radford, I.~Sutskever, and D.~Amodei, ``\BIBforeignlanguage{en}{{Language Models Are Few-Shot Learners}},'' Jul. 2020.

\bibitem{shoeybiMegatronLM2020}
M.~Shoeybi, M.~Patwary, R.~Puri, P.~LeGresley, J.~Casper, and B.~Catanzaro, ``\BIBforeignlanguage{en}{{Megatron-LM: Training Multi-Billion Parameter Language Models Using Model Parallelism}},'' Mar. 2020.

\bibitem{chowdheryPaLM2022}
A.~Chowdhery, S.~Narang, J.~Devlin, M.~Bosma, G.~Mishra, A.~Roberts, P.~Barham, H.~W. Chung, C.~Sutton, S.~Gehrmann, P.~Schuh, K.~Shi, S.~Tsvyashchenko, J.~Maynez, A.~Rao, P.~Barnes, Y.~Tay, N.~Shazeer, V.~Prabhakaran, E.~Reif, N.~Du, B.~Hutchinson, R.~Pope, J.~Bradbury, J.~Austin, M.~Isard, G.~{Gur-Ari}, P.~Yin, T.~Duke, A.~Levskaya, S.~Ghemawat, S.~Dev, H.~Michalewski, X.~Garcia, V.~Misra, K.~Robinson, L.~Fedus, D.~Zhou, D.~Ippolito, D.~Luan, H.~Lim, B.~Zoph, A.~Spiridonov, R.~Sepassi, D.~Dohan, S.~Agrawal, M.~Omernick, A.~M. Dai, T.~S. Pillai, M.~Pellat, A.~Lewkowycz, E.~Moreira, R.~Child, O.~Polozov, K.~Lee, Z.~Zhou, X.~Wang, B.~Saeta, M.~Diaz, O.~Firat, M.~Catasta, J.~Wei, K.~{Meier-Hellstern}, D.~Eck, J.~Dean, S.~Petrov, and N.~Fiedel, ``\BIBforeignlanguage{en}{{PaLM: Scaling Language Modeling with Pathways}},'' Oct. 2022.

\bibitem{Multidatacenter2024}
``\BIBforeignlanguage{en}{{Multi-Datacenter Training: OpenAI's Ambitious Plan to Beat Google's Infrastructure}},'' https://semianalysis.com/2024/09/04/multi-datacenter-training-openais/, Sep. 2024.

\bibitem{mcmahancommunicationefficient2023}
H.~B. McMahan, E.~Moore, D.~Ramage, S.~Hampson, and B.~A. y~Arcas, ``\BIBforeignlanguage{en}{{Communication-Efficient Learning of Deep Networks from Decentralized Data}},'' Jan. 2023.

\bibitem{gorbettCrosssilo2024}
M.~Gorbett, H.~Shirazi, and I.~Ray, ``\BIBforeignlanguage{en}{{Cross-Silo Federated Learning across Divergent Domains with Iterative Parameter Alignment}},'' Apr. 2024.

\bibitem{liangCommunicationefficient2024}
F.~Liang, Z.~Zhang, H.~Lu, V.~C.~M. Leung, Y.~Guo, and X.~Hu, ``\BIBforeignlanguage{en}{{Communication-Efficient Large-Scale Distributed Deep Learning: A Comprehensive Survey}},'' Apr. 2024.

\bibitem{shallueMeasuring2019}
C.~J. Shallue, J.~Lee, J.~Antognini, J.~{Sohl-Dickstein}, R.~Frostig, and G.~E. Dahl, ``\BIBforeignlanguage{en}{{Measuring the Effects of Data Parallelism on Neural Network Training}},'' Jul. 2019.

\bibitem{chenrevisiting2017}
\BIBentryALTinterwordspacing
J.~Chen, X.~Pan, R.~Monga, S.~Bengio, and R.~Jozefowicz. \BIBforeignlanguage{en}{{Revisiting Distributed Synchronous SGD}}. [Online]. Available: \url{http://arxiv.org/abs/1604.00981}
\BIBentrySTDinterwordspacing

\bibitem{douillarddiloco2024}
\BIBentryALTinterwordspacing
A.~Douillard, Q.~Feng, A.~A. Rusu, R.~Chhaparia, Y.~Donchev, A.~Kuncoro, M.~Ranzato, A.~Szlam, and J.~Shen. \BIBforeignlanguage{en-US}{{DiLoCo: Distributed Low-Communication Training of Language Models}}. [Online]. Available: \url{http://arxiv.org/abs/2311.08105}
\BIBentrySTDinterwordspacing

\bibitem{jaghouarOpenDiLoCo2024}
S.~Jaghouar, J.~M. Ong, and J.~Hagemann, ``\BIBforeignlanguage{en}{{OpenDiLoCo: An Open-Source Framework for Globally Distributed Low-Communication Training}},'' Jul. 2024.

\bibitem{douillardstreaming2025}
\BIBentryALTinterwordspacing
A.~Douillard, Y.~Donchev, K.~Rush, S.~Kale, Z.~Charles, Z.~Garrett, G.~Teston, D.~Lacey, R.~McIlroy, J.~Shen, A.~Ramé, A.~Szlam, M.~Ranzato, and P.~Barham. \BIBforeignlanguage{en}{{Streaming DiLoCo with Overlapping Communication: Towards a Distributed Free Lunch}}. [Online]. Available: \url{http://arxiv.org/abs/2501.18512}
\BIBentrySTDinterwordspacing

\bibitem{narayananPipeDream2019}
D.~Narayanan, A.~Harlap, A.~Phanishayee, V.~Seshadri, N.~R. Devanur, G.~R. Ganger, P.~B. Gibbons, and M.~Zaharia, ``\BIBforeignlanguage{en}{{PipeDream: Generalized Pipeline Parallelism for DNN Training}},'' in \emph{\BIBforeignlanguage{en}{Proceedings of the 27th ACM Symposium on Operating Systems Principles}}.\hskip 1em plus 0.5em minus 0.4em\relax Huntsville Ontario Canada: ACM, Oct. 2019, pp. 1--15.

\bibitem{chenSAPipe2022}
Y.~Chen, C.~Xie, M.~Ma, J.~Gu, Y.~Peng, H.~Lin, C.~Wu, and Y.~Zhu, ``\BIBforeignlanguage{en}{{SAPipe: Staleness-Aware Pipeline for Data Parallel DNN Training}},'' \emph{\BIBforeignlanguage{en}{Advances in Neural Information Processing Systems}}, vol.~35, pp. 17\,981--17\,993, Dec. 2022.

\bibitem{narayananEfficient2021}
D.~Narayanan, M.~Shoeybi, J.~Casper, P.~LeGresley, M.~Patwary, V.~A. Korthikanti, D.~Vainbrand, P.~Kashinkunti, J.~Bernauer, B.~Catanzaro, A.~Phanishayee, and M.~Zaharia, ``\BIBforeignlanguage{en}{{Efficient Large-Scale Language Model Training on GPU Clusters Using Megatron-LM}},'' Aug. 2021.

\bibitem{liStragglers2021}
X.~Li, Z.~Qu, B.~Tang, and Z.~Lu, ``\BIBforeignlanguage{en}{Stragglers are not disaster: A hybrid federated learning algorithm with delayed gradients},'' Feb. 2021.

\bibitem{bischofStructured1993}
C.~Bischof, C.~, G., and A.~{and Griewank}, ``\BIBforeignlanguage{en}{Structured second-and higher-order derivatives through univariate taylor series},'' \emph{\BIBforeignlanguage{en}{Optimization Methods and Software}}, vol.~2, no. 3-4, pp. 211--232, Jan. 1993.

\bibitem{zhengAsynchronous2020}
S.~Zheng, Q.~Meng, T.~Wang, W.~Chen, N.~Yu, Z.-M. Ma, and T.-Y. Liu, ``\BIBforeignlanguage{en}{{Asynchronous Stochastic Gradient Descent with Delay Compensation}},'' Feb. 2020.

\bibitem{rigazziDCS3GD2019}
A.~Rigazzi, ``\BIBforeignlanguage{en}{{DC-S3GD: Delay-Compensated Stale-Synchronous SGD for Large-Scale Decentralized Neural Network Training}},'' in \emph{\BIBforeignlanguage{en}{2019 IEEE/ACM Third Workshop on Deep Learning on Supercomputers (DLS)}}.\hskip 1em plus 0.5em minus 0.4em\relax Denver, CO, USA: IEEE, Nov. 2019, pp. 62--68.

\bibitem{wangAsynchronous2023}
H.~Wang, Z.~Jiang, C.~Liu, S.~Sarkar, D.~Jiang, and Y.~M. Lee, ``\BIBforeignlanguage{en}{{Asynchronous Training Schemes in Distributed Learning with Time Delay}},'' \emph{\BIBforeignlanguage{en}{Transactions on Machine Learning Research}}, Nov. 2023.

\bibitem{friedman2001elements}
J.~Friedman, T.~Hastie, and R.~Tibshirani, ``{The elements of statistical learning. vol. 1 Springer series in statistics},'' \emph{New York}, 2001.

\bibitem{amariNatural1998}
S.-I. Amari, ``\BIBforeignlanguage{en}{{Natural Gradient Works Efficiently in Learning}},'' \emph{\BIBforeignlanguage{en}{Neural Computation}}, vol.~10, no.~2, pp. 251--276, 1998.

\bibitem{liPyTorch2020}
S.~Li, Y.~Zhao, R.~Varma, O.~Salpekar, P.~Noordhuis, T.~Li, A.~Paszke, J.~Smith, B.~Vaughan, P.~Damania, and S.~Chintala, ``\BIBforeignlanguage{en}{Pytorch distributed: Experiences on accelerating data parallel training},'' Jun. 2020.

\bibitem{touvronLLaMA2023}
H.~Touvron, T.~Lavril, G.~Izacard, X.~Martinet, M.-A. Lachaux, T.~Lacroix, B.~Rozi{\`e}re, N.~Goyal, E.~Hambro, F.~Azhar, A.~Rodriguez, A.~Joulin, E.~Grave, and G.~Lample, ``\BIBforeignlanguage{en}{Llama: Open and efficient foundation language models},'' Feb. 2023.

\bibitem{raffelExploring2023}
C.~Raffel, N.~Shazeer, A.~Roberts, K.~Lee, S.~Narang, M.~Matena, Y.~Zhou, W.~Li, and P.~J. Liu, ``\BIBforeignlanguage{en}{Exploring the limits of transfer learning with a unified text-to-text transformer},'' Sep. 2023.

\bibitem{kalamkarStudy2019}
D.~Kalamkar, D.~Mudigere, N.~Mellempudi, D.~Das, K.~Banerjee, S.~Avancha, D.~T. Vooturi, N.~Jammalamadaka, J.~Huang, H.~Yuen, J.~Yang, J.~Park, A.~Heinecke, E.~Georganas, S.~Srinivasan, A.~Kundu, M.~Smelyanskiy, B.~Kaul, and P.~Dubey, ``\BIBforeignlanguage{en}{A study of bfloat16 for deep learning training},'' Jun. 2019.

\end{thebibliography}

% \newpage
% \appendices

% \section{Convergence Proof of PACT}\label{appendix1}
% \input{appendix.tex}

\end{document}